\documentstyle[12pt,epsf,epsfig]{article}
\newcount\timecount
\newcount\hours \newcount\minutes  \newcount\temp \newcount\pmhours

\hours = \time
\divide\hours by 60
\temp = \hours
\multiply\temp by 60
\minutes = \time
\advance\minutes by -\temp
\def\hour{\the\hours}
\def\minute{\ifnum\minutes<10 0\the\minutes
            \else\the\minutes\fi}
\def\clock{
\ifnum\hours=0 12:\minute\ AM
\else\ifnum\hours<12 \hour:\minute\ AM
      \else\ifnum\hours=12 12:\minute\ PM
            \else\ifnum\hours>12
                 \pmhours=\hours
                 \advance\pmhours by -12
                 \the\pmhours:\minute\ PM
                 \fi
            \fi
      \fi
\fi
}

\def\monthname{\relax\ifcase\month 0/\or January\or February\or
   March\or April\or May\or June\or July\or August\or September\or
   October\or November\or December\else\number\month/\fi}

\def\bold#1{\setbox0=\hbox{$#1$}%
     \kern-.025em\copy0\kern-\wd0
     \kern.05em\copy0\kern-\wd0
     \kern-.025em\raise.0433em\box0 }

\def\gappeq{\mathrel{\rlap {\raise.5ex\hbox{$>$}}
{\lower.5ex\hbox{$\sim$}}}}

\def\lappeq{\mathrel{\rlap{\raise.5ex\hbox{$<$}}
{\lower.5ex\hbox{$\sim$}}}}

\def\ga{\mathrel{\raise.3ex\hbox{$>$\kern-.75em\lower1ex\hbox{$\sim$}}}}
\def\la{\mathrel{\raise.3ex\hbox{$<$\kern-.75em\lower1ex\hbox{$\sim$}}}}
\def\gev{{\rm \, Ge\kern-0.125em V}}
\def\tev{{\rm \, Te\kern-0.125em V}}
\def\beq{\begin{equation}}
\def\eeq{\end{equation}}

\def\ohsq{\Omega_{\chi} h^2}

\def\m12{m_{1\!/2}}

\begin{document}
\begin{titlepage}
\pagestyle{empty}
\baselineskip=21pt
\rightline{hep-ph/0102098}
\rightline{CERN--TH/2001-024, ROM2F/2001/05}
\rightline{UMN--TH--1938/01, TPI--MINN--01/07}
\vskip 0.05in
\begin{center}
{\large{\bf
The CMSSM Parameter Space at Large $\tan\beta$}}
\end{center}
\begin{center}
\vskip 0.05in
{{\bf John Ellis}$^1$, 
{\bf Toby Falk}$^2$,
{\bf Gerardo Ganis}$^3$,
{\bf Keith A.~Olive}$^{1,2}$ and
{\bf Mark Srednicki}$^4$}\\
\vskip 0.05in
{\it
$^1${TH Division, CERN, Geneva, Switzerland}\\
$^2${Theoretical Physics Institute, School of Physics and Astronomy,\\
University of Minnesota, Minneapolis, MN 55455, USA}\\
$^3${INFN, Universit\`a di Roma II `Tor Vergata', Rome, Italy}\\
$^4${Department of Physics, University of California, Santa Barbara,
CA~93106, USA}\\
}
\vskip 0.05in
{\bf Abstract}
\end{center}
\baselineskip=18pt \noindent

We extend previous combinations of LEP and cosmological relic density
constraints on the parameter space of the constrained MSSM, with universal
input supersymmetry-breaking parameters, to large $\tan \beta$.  We take
account of the possibility that the lightest Higgs boson might weigh about
115~GeV, but also retain the possibility that it might be heavier.  We
include the most recent implementation of the $b \to s \gamma$ constraint
at large $\tan \beta$.  We refine previous relic density calculations at
large $\tan \beta$ by combining a careful treatment of the direct-channel
Higgs poles in annihilation of pairs of neutralinos $\chi$ with a complete
treatment of $\chi - {\tilde \tau}$ coannihilation, and discuss carefully
uncertainties associated with the mass of the $b$ quark. We find that
coannihilation and pole annihilations allow the CMSSM to yield an
acceptable relic density at large $\tan \beta$, but it is consistent with
all the constraints only if $m_\chi > 140 ~ (180)$~GeV for $\mu > 0 ~ (\mu
< 0)$ for our default choices $m_b(m_b)^{\overline {MS}}_{SM} = 4.25$ GeV,
$m_t = 175$ GeV, and $A_0 = 0$.

\vfill
\vskip 0.15in
\leftline{CERN--TH/2001-024}
\leftline{February 2001}
\end{titlepage}
\baselineskip=18pt

Now that LEP has been terminated, and future accelerator constraints on
supersymmetry may take a while to accumulate, it is important to extract
the last drop of phenomenological information from the completed LEP
searches~\cite{Junk,LEPHiggs}. The constraints from LEP are particularly
interesting when combined with the measured value of the $b \to s \gamma$
decay rate~\cite{bsgexpt} and with restrictions on cold dark matter
imposed by astrophysics and cosmology, assuming that the lightest
supersymmetric particle (LSP) is the lightest neutralino
$\chi$~\cite{EHNOS}, and that $R$ parity is conserved. We~\cite{EFOS,
EFOSi,EFGO,EGNO} and others~\cite{others,dn,BB,eg,BK,glp,LNS} have made
such
combinations, in both the generic minimal supersymmetric extension of the
Standard Model (MSSM) and with the supplementary constraint that the soft
supersymmetry-breaking scalar masses be universal at some high input scale
(CMSSM), as in minimal supergravity (mSUGRA)  models~\cite{reviews}. 
We limited our previous analyses~\cite{EFOS,EFOSi,EGNO} to
$\tan
\beta
\le 20$, because the available $b \to s \gamma$ calculations~\cite{bsgcalx}
were not applicable to larger values of $\tan \beta$, and because we were
dissatisfied with the reliability and accuracy of the available
calculations at larger $\tan \beta$ of the the relic density $\ohsq$, for
which we regard $0.1\leq\ohsq\leq 0.3$ as the cosmologically-favoured
range. 

Concerning the first point, we note that new and improved calculations for
large $\tan \beta$ have recently come available~\cite{newbsgcalx}, and we
implement them in our analysis.
Concerning $\ohsq$, we note two important issues. The
first is that $\chi - {\tilde \ell}$ coannihilation effects are important
in the CMSSM~\cite{EFOSi}, and the second is that direct-channel
annihilations through poles~\cite{directpoles} due to the heavier neutral
MSSM Higgs bosons $H$ and $A$
are of increasing importance for larger $\tan
\beta$~\cite{dn,BB,lns}.

We recall that, in the CMSSM, the next-to-lightest supersymmetric particle
(NLSP) is the lighter stau ${\tilde \tau}_1$ in a generic domain of
parameter space, and $\chi - {\tilde \tau}, {\tilde \mu}$ and ${\tilde e}$
coannihilations are important for calculating the relic
density~\footnote{In the CMSSM, neither $\chi - \chi' -
\chi^\pm$~\cite{MY,eg} nor $\chi - {\tilde t}$~\cite{BD} coannihilations
are important.}. The most complete published $\chi - {\tilde \ell}$
coannihilation calculations known to us are those in~\cite{EFOSi,glp}, but
more $\chi - {\tilde \tau}$ coannihilation diagrams become potentially
relevant at large $\tan \beta$. 

As for the $A, H$ poles, we recall that,
unlike the lightest MSSM Higgs boson $h$, the masses and total decay
widths of the $H$ and $A$ increase with the soft supersymmetry-breaking
parameters $m_0, m_{1/2}$, enhancing their significance. An
accurate treatment of the bottom-quark mass $m_b$, its renormalization and
that of the corresponding Yukawa coupling is important for the correct
calculation of the Higgs spectrum and couplings at large $\tan \beta$.
Here we include with similar accuracy the corresponding
effects associated with the mass and Yukawa coupling of the $\tau$ lepton. 
Calculations of $\chi - \chi$ annihilations via a series
expansion in $x_f$, the freeze-out temperature $T_f$ divided by $m_\chi$,
are inadequate when $m_{H,A} \sim 2 m_\chi$ and the $H$ and $A$ poles are
important.  Calculations of $\chi - \chi$ annihilations that go beyond the
$x_f$ series expansion are available~\cite{BB,gond}, but they have not yet
been combined with complete coannihilation calculations at large $\tan
\beta$. As we see later, there is an important interplay between
annihilations via the direct-channel $A,H$ poles and coannihilation
processes. 

We find that the CMSSM survives all the experimental and cosmological
constraints at
large $\tan \beta$ only thanks to the coannihilation and $A, H$ pole
annihilation effects. Combining the available constraints, we find
that $m_\chi \ga 140 ~(180)$~GeV for $\mu > 0 ~ (\mu < 0)$ and our default
choices $m_t = 175$~GeV, $m_b(m_b)^{\overline {MS}}_{SM} = 4.25$~GeV, 
and $A_0 = 0$. We also find an upper bound $m_\chi \la 400$ to 550~GeV
for $\tan \beta \ge 20$, {\it if} $m_h \sim 115$~GeV, as suggested by
LEP~\cite{LEPHiggs}.

The most important experimental constraints on the CMSSM parameter space
are provided by LEP searches for sparticles and Higgs bosons~\cite{Junk},
the latter constraining the sparticle spectrum indirectly via radiative
corrections. The kinematic reach for charginos was $m_{\chi^\pm} =
104$~GeV, and the LEP limit is generally close to this value, within the
CMSSM framework. The LEP searches for sleptons impose $m_{\tilde e} >
97$~GeV,
$m_{\tilde \mu} > 94$~GeV and $m_{\tilde \tau} > 80$~GeV for $m_\chi <
80$~GeV. The only one of these to fall noticeably below the limit expected
statistically is that on the ${\tilde \tau}$, but this is not interpreted
as significant evidence for a ${\tilde \tau}$. Other important sparticle
constraints are those on stop squarks ${\tilde t}$: $m_{\tilde t} >
94$~GeV for $m_\chi < 80$~GeV from LEP, and $m_{\tilde t} \ga 115$~GeV
for
$m_\chi \la 50$~GeV from the Fermilab Tevatron collider~\cite{stopT}.

The lower limit on the mass of a Standard Model Higgs boson imposed by the
combined LEP experiments
is $113.5$~GeV~\cite{LEPHiggs}. This lower limit applies also to the MSSM
for small $\tan \beta$, even if squark mixing is maximal. In the CMSSM,
maximal mixing is not attained, and the $e^+ e^- \to Z^0 + h$ production
rate is very similar to that in the Standard Model~\cite{ZH}, for all
values of $\tan \beta$. As is well known, a 2.9-$\sigma$ signal for a
Higgs boson weighing about $115^{+1.3}_{-0.7}$~GeV has been
reported~\cite{LEPHiggs}. At points in the following, we comment
explicitly on the possible implications if $m_h \sim 115~{\rm
GeV}$~\cite{EGNO}. In
order to account for uncertainties in theoretical calculations of $m_h$ in
the MSSM~\cite{HHH} for any given value of $m_t$, we consider this
LEP range~\cite{LEPHiggs} to be consistent with CMSSM parameter
choices that yield $113~{\rm GeV} \le m_h \le
117~{\rm GeV}$. The theoretical value of $m_h$ in the MSSM is quite
sensitive to $m_t$, the pole mass of the top quark: we use $m_t = 175$~GeV
as default, but also discuss
briefly the cases $m_t = 170, 180$~GeV. 

We implement the new NLO $b \to s \gamma$ calculations
of~\cite{newbsgcalx} when ${\tilde M} > 500$~GeV, where ${\tilde M} = {\rm
Min}(m_{\tilde q}, m_{\tilde g})$. Otherwise, we use only the LO
calculations and assign a larger theoretical error. For the experimental
value, we combine the CLEO measurement with the recent BELLE
result~\cite{bsgexpt}, assuming full correlation between the experimental
systematics~\footnote{This is conservative, but the available information
does not justify a more restrictive approach, and this assumption is in
any case not very important.}, finding ${\cal B} (b \to s \gamma) = (3.21
\pm 0.44 \pm 0.26) \times 10^{-4}$. In our implementation, we allow CMSSM
parameter choices that, after including the theoretical errors due to the
scale and model dependences, may fall within the 95\% confidence level
range $2.33 \times 10^{-4} < {\cal B}(b \to s < \gamma) < 4.15 \times
10^{-4}$. In general, we find in the regions excluded when $\mu < 0$ that
the predicted value of ${\cal B} (b \to s \gamma)$ is larger than this
measured range, whereas, when $\mu > 0$, the exclusion results from ${\cal
B} (b \to s \gamma)$ being smaller than measured.

There is a tendency for the masses of
the (nearly degenerate) $H$ and $A$ bosons to drop at large $\tan \beta$
in the CMSSM, and hence for strong direct-channel annihilation: $\chi \chi
\to H, A
\to X$ become possible when $m_{H,A} \sim 2 m_\chi$.  
Annihilation via the
$A$ pole to ${\bar b} b$ is additionally enhanced because of the large
$A {\bar b} b$ coupling at large $\tan \beta$.  
The $H {\bar b} b$ coupling is equally enhanced,
but the $A$
pole is more important because direct-channel pseudoscalar exchange to
${\bar f} f$ final states is not $P$-wave suppressed.
In view of the importance of the direct-channel $A$ pole at large $\tan
\beta$, we take pains to include the most accurate available calculation
of $m_A$, which incorporates a renormalization-group (RG) improvement of
the standard effective potential calculation, so as to take into account
the leading effects associated with the third-generation Yukawa
couplings~\footnote{It might also be desirable to include the
corresponding improvement linked to gauge couplings~\cite{KLNS}, but we
do not have these available, and they are likely to be less important.}. 
This calculation may be extracted from~\cite{CEPW}, setting to zero the
CP-violating phases. We parametrize the tree-level MSSM Higgs potential in
the usual way, denoting by $m_{12}^2$ the coefficient of the
$H^{\dagger}_1 H_2$ term in the effective lagrangian, and include one-loop
corrections due to the $\tau$
lepton and the ${\tilde \tau}$ sleptons and the $b$ supermultiplet - in
view of their potential importance at large $\tan \beta$ - as well as the
$t$ supermultiplet. The RG-improved pseudoscalar MSSM Higgs mass is then
given by
\begin{equation}
m_A^2 = {{\rm Re} m_{12}^2 + {\rm Re} {m}_{12}^{2(1)} \over \sin \beta
\cos \beta} + \Sigma_{f = t, b, \tau} {(\Delta {\cal M}_A^2)^{\tilde f}
\over
\xi^{\tilde f}_1(m_t) \xi^{\tilde f}_2(m_t)}
\label{CEPW}
\end{equation}
where ${\rm Re} {m}_{12}^{2(1)}$ is taken from equation (3.7) 
of~\cite{CEPW}, $(\Delta {\cal M}_A^2)^{\tilde f}$ is the standard 
one-loop
contribution to $m_A^2$ due to the third-generation fermions $f$, and
$\xi^{\tilde f}_{1(2)}(m_t)$ are anomalous-dimension factors taken from
equation (3.25) of~\cite{CEPW}~\footnote{On the other hand, we do not
include the corresponding RG improvements to $m_{h,H,H^\pm}$~\cite{CEPW},
which are more complicated but less essential for our purposes. The
effects of the direct-channel $H$ pole are hidden by the more important
effects of the almost degenerate $A$ pole.}. 

It is well known
that the partial-wave expansion breaks down in the vicinity of
poles~\cite{directpoles}, whose finite widths must be
taken into account. We calculate these including all the SM final
states and relevant QCD corrections \cite{Spira}.  To account for
the poles, we perform the full phase-space integration~\cite{BB} for the
direct-s channel $A$ and
$H$
exchanges in the vicinities of their poles, namely for $0.65 < 2
m_{\chi}/m_{A,H} < 2.0$. This full phase-space integration must be
performed for temperatures $T$ down to the freeze-out temperature $T_f =
x_f \times m_\chi$.  When the direct-channel poles are important, we
determine the correct freeze-out temperature iteratively.

As already mentioned, $\chi - {\tilde \ell}$ coannihilation - particularly
that
with the lighter stau ${\tilde \tau}_1$ - is important in the CMSSM,
extending significantly the allowed range of $m_{1/2}$.  The most complete
available calculations of $\chi - {\tilde \ell}$
coannihilation~\cite{EFOSi} are, however, inadequate at large $\tan
\beta$, for example because the larger $\tau$ Yukawa coupling in this
limit increases the importance of diagrams that were negligible for
smaller $\tan \beta$.  In updating our coannihilation code, we have also
taken the opportunity to complete and improve some of its other aspects.
The following are the most important modifications: (a) ${\tilde \tau}_1 -
{\tilde \tau}_2$ mixing has been incorporated fully (previously, it was
incorporated in the kinematics but not in the couplings), (b) $m_\tau$
effects have been included in ${\tilde \tau} - {\tilde e}, {\tilde \mu}$
coannihilations, (c) crossed-channel ${\tilde \tau}_2$-exchange diagrams
have been included for coannihilations into the following final states: 
$Z^0 Z^0, Z^0 h, h h, \tau Z^0, \tau h$, and (d) subroutines for the
following final states:  $Z^0 H, \gamma H, h A, H A, W^\pm H^\mp, A A, h
H, H H, H^+ H^-, \tau H$ and $\tau A$ have been included in the code
used to obtain the results we present here~\footnote{Details of these
calculations will appear in~\cite{EFGOSi2}.}. Results for some
of these processes were previously presented in~\cite{EFOSi}, but
they were kinematically inaccessible in the region of interest there.
We find that for $\tan \beta
\le 20$ inclusion of the new coannihilation effects does not alter
significantly the region of the $(m_{1/2}, m_0)$ plane favoured by
cosmology, whereas for $\tan \beta \ge 30$ there is a significant increase
of the favoured range of $m_{1/2}$ in the coannihilation region, as we
discuss in more detail later.  

Since the precise value of the $b$-quark mass $m_b$ is important at large
$\tan \beta$, we now discuss it in some detail. Various different
definitions of $m_b$ have been proposed, and we prefer the running mass in
the ${\overline {MS}}$ renormalization prescription for the Standard Model
(SM): $m_b(m_b)^{\overline {MS}}_{SM}$. A recent determination of
$m_b(m_b)^{\overline {MS}}_{SM}$ from a combination of lattice
calculations and experimental measurements of the masses of $B$
mesons~\cite{MS} yields $m_b(m_b)^{\overline {MS}}_{SM} = 4.25 \pm
0.15$~GeV~\footnote{This also includes the ranges favoured by recent
sum-rule analyses of $\Upsilon$ masses~\cite{Upsilon}.}. We therefore take
as our default value $m_b(m_b)^{\overline {MS}}_{SM} = 4.25$~GeV, but also
discuss later the change in physics if $m_b(m_b)^{\overline {MS}}_{SM} =
4.0$ or 4.5~GeV, regarding this as a full $\pm 2 - \sigma$ range. We note
that this is similar to the range quoted by the Particle Data
Group~\cite{PDG}. Having
fixed the input value, we then evolve the running
mass $m_b(Q)^{\overline {MS}}_{SM}$ from $Q = m_b$ up to $Q = m_Z$ using
the three-loop ${\overline {MS}}$ RGE's for the Standard
Model~\cite{threeloopmb}
\beq
m_b(m_Z)^{\overline {MS}}_{SM} = m_b(m_b)^{\overline
{MS}}_{SM}\left[{\alpha_s(m_Z) \over \alpha_s(m_b)}
\right]^{12/23} {c[\alpha_s(m_Z)/\pi] \over c[\alpha_s(m_b)/\pi]}
\eeq
where 
\beq
c[\alpha_s] = 1 + 1.175 \alpha_s + 1.501 \alpha_s^2 - 0.172 \alpha_s^3
\eeq
which we evaluate using $\alpha_s(4.25~{\rm GeV}) = 0.2246$ and
$\alpha_s(m_Z) =
0.1185$. We then convert $m_b$ to the ${\overline {DR}}$ scheme, using
the one-loop correction factor 
\beq
m_b(m_Z)^{\overline
{DR}}_{SM} = m_b(m_Z)^{\overline {MS}}_{SM} \times [1 - (\alpha_s /3 \pi)
+ (3 \alpha_2 / 32 \pi)]
\eeq
valid in the Standard Model~\cite{MStoDR}. We
then make the further correction to convert from the Standard Model value
$m_b(m_Z)^{\overline {DR}}_{SM}$ to the MSSM value $m_b(m_Z)^{\overline
{DR}}_{MSSM}$~\cite{SMtoMSSM}. Finally, we use the MSSM ${\overline {DR}}$
RGE's at scales between $m_Z$ and the unification scale~\cite{RGEs}. We
use the running $m_{b,t} (Q = 2 m_\chi)$ when evaluating annihilation
processes.

\begin{figure}
\vspace*{-0.75in}
\hspace*{-.70in}
\begin{minipage}{8in}
\epsfig{file=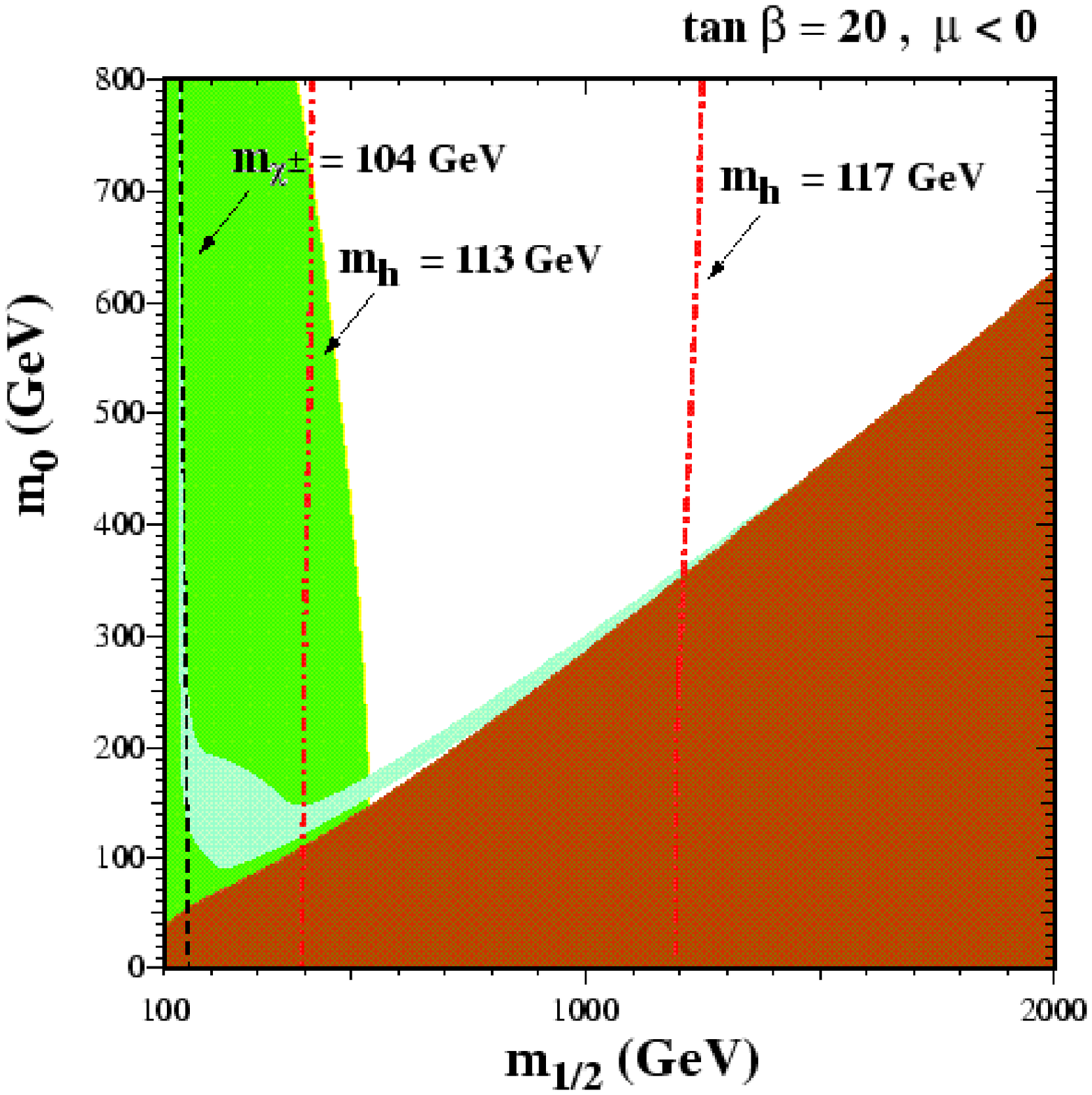,height=3.5in}
\hspace*{-0.17in}
\epsfig{file=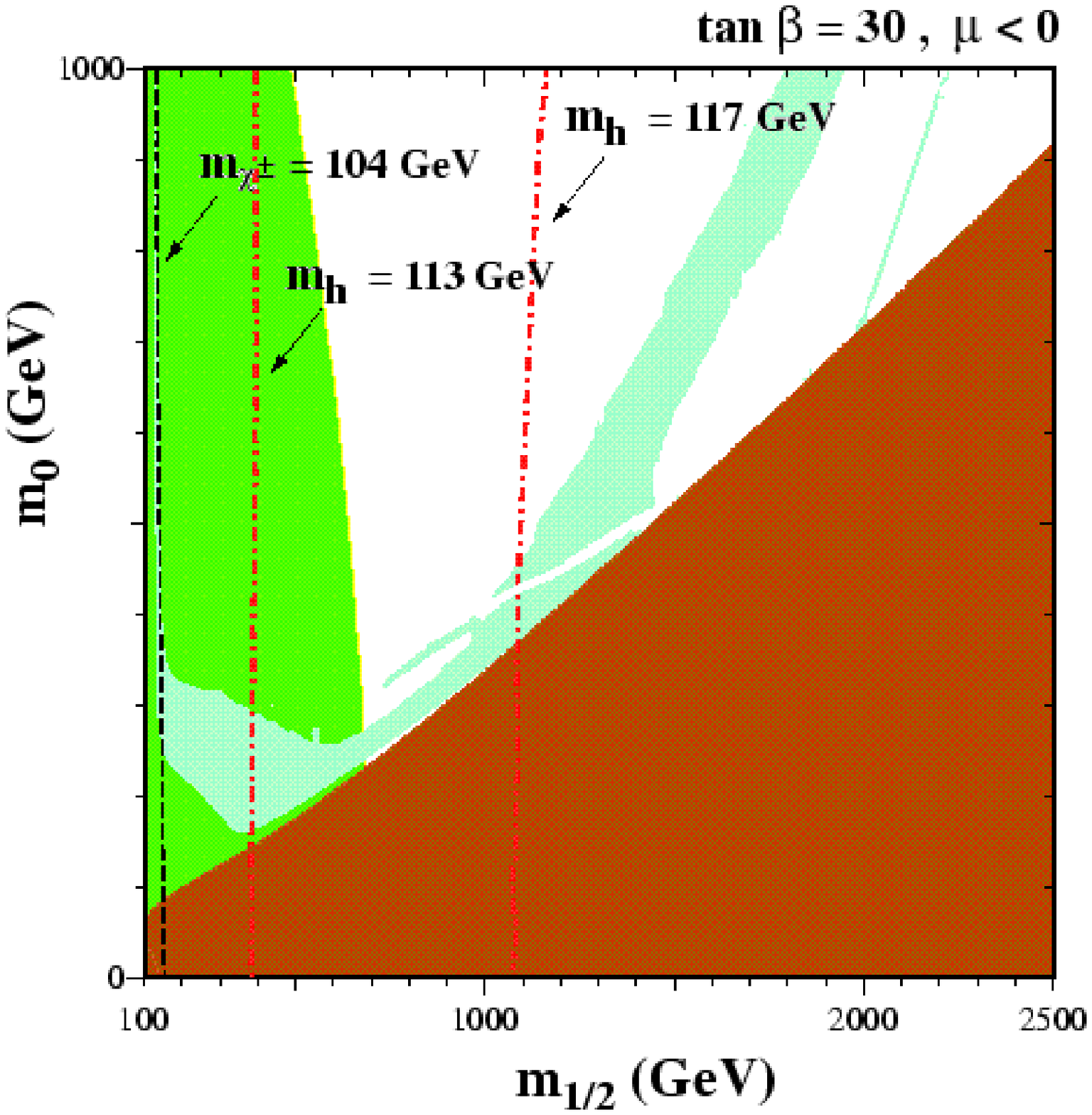,height=3.5in} \hfill
\end{minipage}
\hspace*{-.70in}
\begin{minipage}{8in}
\epsfig{file=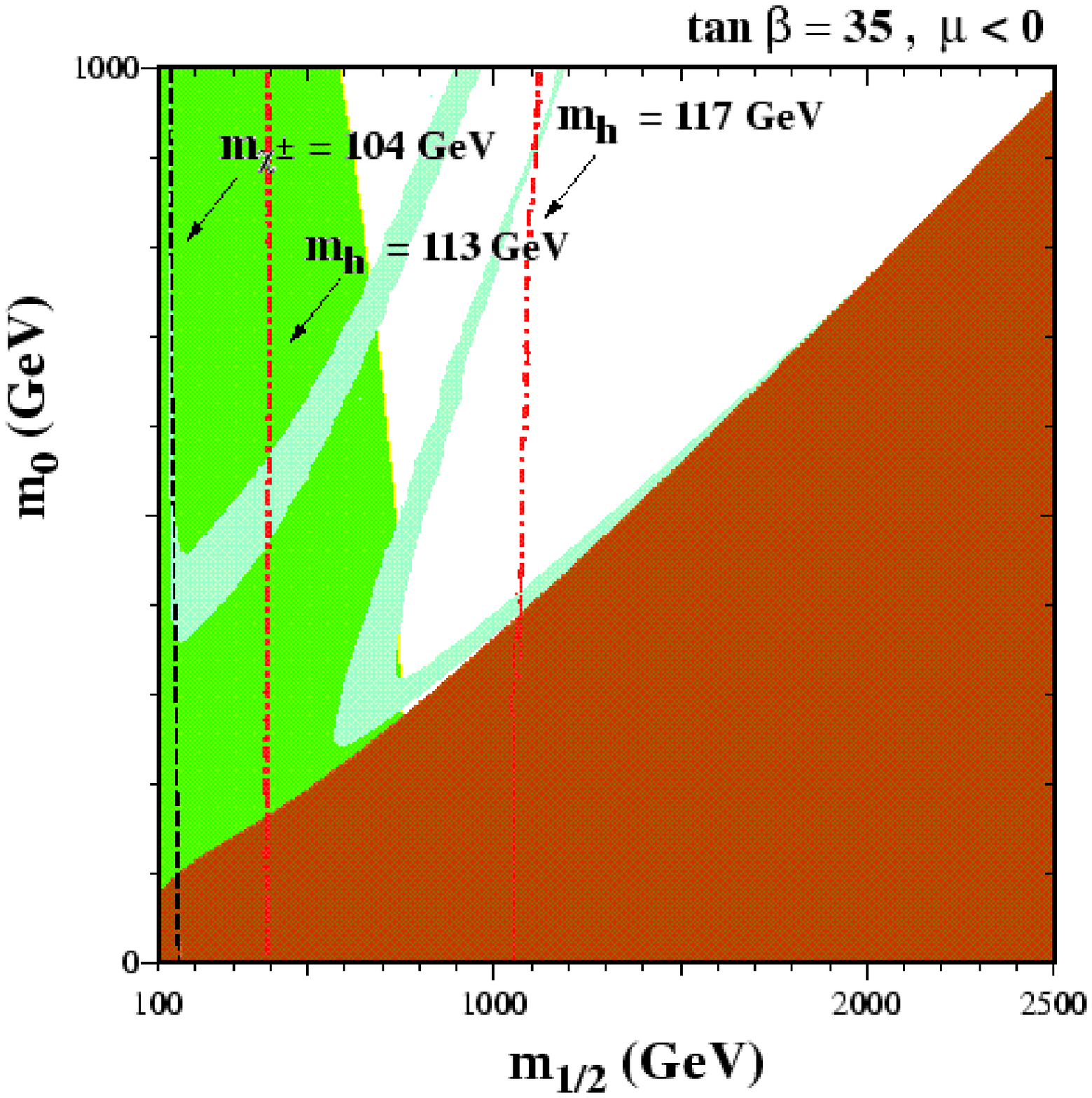,height=3.5in}
\hspace*{-0.2in}
\epsfig{file=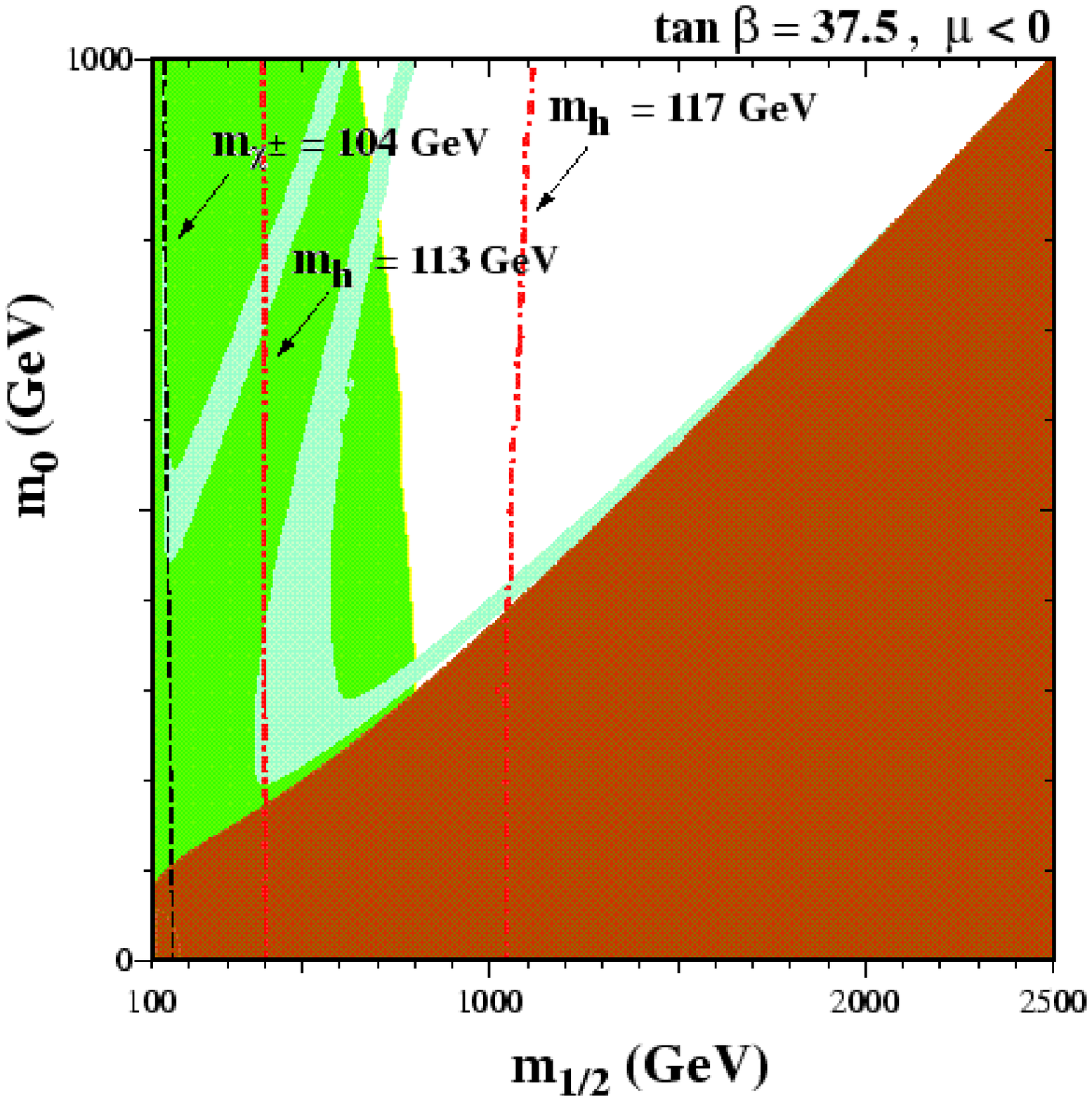,height=3.5in} \hfill
\end{minipage}
\caption{\label{fig:negative}
{\it The $(m_{1/2}, m_0)$ planes for $\mu < 0$ and $\tan \beta =$ (a) 20,
(b) 30, (c) 35 and (d) 37.5, found assuming $A_0 = 0, m_t = 175$~GeV and
$m_b(m_b)^{\overline {MS}}_{SM} = 4.25$~GeV. In this case, we find no
large allowed region for
$\tan \beta \ge 40$. The near-vertical are 
 the contours
$m_{\chi^\pm} = 104$~GeV (dashed), $m_h = 113, 117$~GeV (dot-dashed). 
The medium (dark green) shaded regions are excluded by $b
\to s \gamma$.
The light (turquoise) shaded areas are the cosmologically
preferred
regions with \protect\mbox{$0.1\leq\ohsq\leq 0.3$}. Away from the pole,
above (below) these light-shaded areas, the relic density $\ohsq > 0.3 (<
0.1)$. In the
dark (brick red) shaded regions, the LSP is the charged ${\tilde \tau}_1$,
so
this
region is excluded. 
The diagonal channel
of low relic densities visible for $\tan \beta \ge 30$, flanked on both
sides by cosmologically preferred regions, is due to direct-channel
annihilation via the $A, H$ poles.}}
\end{figure}

Since it is of interest for most sparticle searches, and for ease of
comparison with our previous results, we first present results in the
$(m_{1/2}, m_0)$ plane. For definiteness, as default we choose the
trilinear soft supersymmetry-breaking parameter $A_0 = 0$ at the input GUT
scale, and determine $\mu$ (up to a sign ambiguity) from the electroweak
vacuum conditions for the specified value of $\tan \beta$. We also use as
default $m_t = 175$~GeV for the on-shell top-quark mass. With our default
choices, and if $\mu < 0$, there are no large regions with consistent
electroweak vacua for $\tan \beta \ge 40$, and we do not expect our new
results to differ significantly from our previous results~\cite{EFOSi} for
$\tan \beta < 20$. We display in Fig.~\ref{fig:negative} the $(m_{1/2},
m_0)$ planes for $\mu < 0$ and $\tan \beta = 20, 30, 35$ and $37.5$. We
see that the LEP Higgs constraint and $b \to s \gamma$ each exclude large
regions with small $m_{1/2}$: the only supersymmetric dark 
matter regions to survive for $\mu < 0$ are those made
possible by $\chi - {\tilde \ell}$ coannihilation and/or rapid
annihilations via the $A, H$ poles.

We see in panel (a) that our new results are indeed very similar to our
previous results~\cite{EFOSi} for $\tan \beta = 20$. The only (small)
differences are due to the improved treatment of $m_b$ discussed above,
which is relevant in the region $m_0 \sim m_{1/2} \sim
200$~GeV, and at the boundary of the coannihilation region.  On the other
hand, we see a dramatic new feature in panels (b), (c) and (d) of
Fig.~\ref{fig:negative} for $\tan \beta \ge 30$: rapid annihilation
through the direct-channel $A, H$ poles suppresses the relic density:
$\Omega_\chi h^2 < 0.1$ along a steep diagonal strip in the $(m_{1/2},
m_0)$ plane where $m_\chi \sim m_A/2$. This is flanked by two allowed
bands where $0.1 < \Omega_\chi h^2 < 0.3$, that connect with the $\chi -
{\tilde \tau}$ coannihilation region on one side and with the
low-$m_{1/2}$ region on the other side. This feature develops
first at a relatively large ratio of $m_{1/2} / m_0$ for $\tan \beta =
30$, and that the ratio decreases as $\tan \beta$
increases~\footnote{Conversely, for smaller values of $\tan \beta$, the
$A,H$-annihilation strip appears at larger $m_{1/2}$, not intersecting
the cosmologically-preferred region identified previously~\cite{EFOSi},
and the direct-channel $A,H$ pole is not important.}. Although difficult
to see,
there is a very narrow allowed band to the right of the
$A,H$-annihilation strip in panel (b)  for $\tan \beta = 30$ with
width $\delta m_{1/2} \sim 15$~GeV, whereas the corresponding band to the
left
of the $A,H$-annihilation strip is clearly visible, with a larger width
$\delta m_{1/2} \sim
150$~GeV.
The width of the coannihilation strip that appears when $m_{{\tilde
\tau}_1} \ga m_\chi$ has a width $\delta m_0 \sim 30$~GeV, which does not 
depend much on $\tan \beta$. However, as seen in
panels (c) and (d), the relative widths of the allowed bands on either
side of the $A,H$-annihilation strip change as $\tan \beta$
increases~\footnote{We do not discuss here in detail the `focus point'
region where an acceptable relic density may be obtained in a band of
thickness $\delta m_0 \sim 30$~GeV when $m_0 \sim 3$~TeV~\cite{Feng}. We
note that our results on $A, H$ pole effects at large $\tan \beta$
differ from~\cite{Feng}, apparently because of a different treatment of
$m_b$, in particular.}. 

We also notice in panel (b) another narrow near-horizontal band of
suppressed
relic density, which meets the first when $m_{1/2} \sim 1500$~GeV and $m_0
\sim 500$~GeV, and is due to rapid ${\tilde \tau}_1 {\bar {\tilde
\tau}_1}
\to H$ annihilation. This not only suppresses the relic density: $\ohsq
< 0.1$ in a band crossing the
left flank of the $\chi \chi \to A, H$ annihilation strip, but also
suppresses the relic density along a narrow band at lower $m_{1/2}$,
reducing $\ohsq$ into the allowed range.

\begin{figure}
\vspace*{-0.75in}
\hspace*{-.70in}
\begin{minipage}{8in}
\epsfig{file=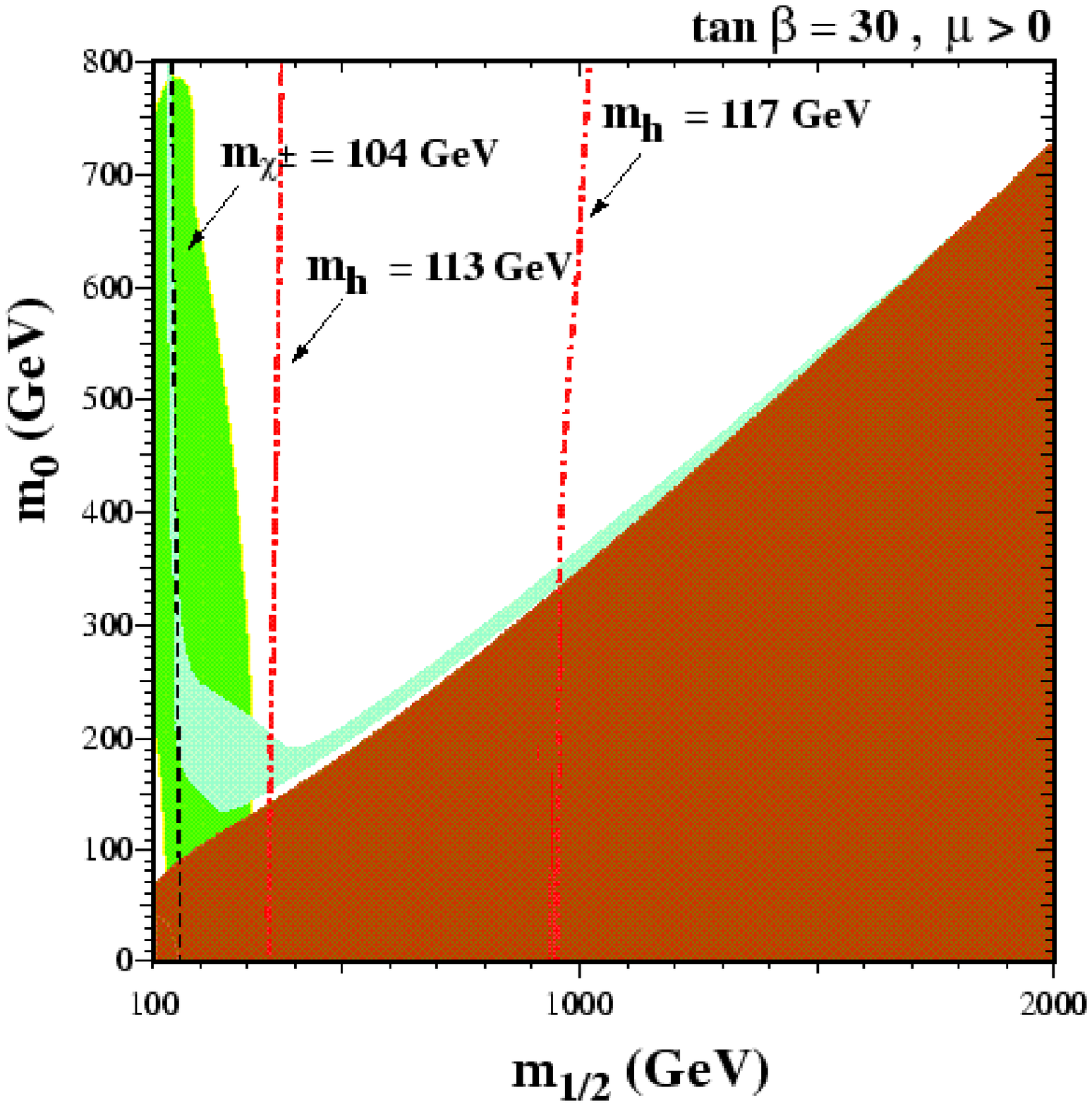,height=3.5in}
\hspace*{-0.15in}
\epsfig{file=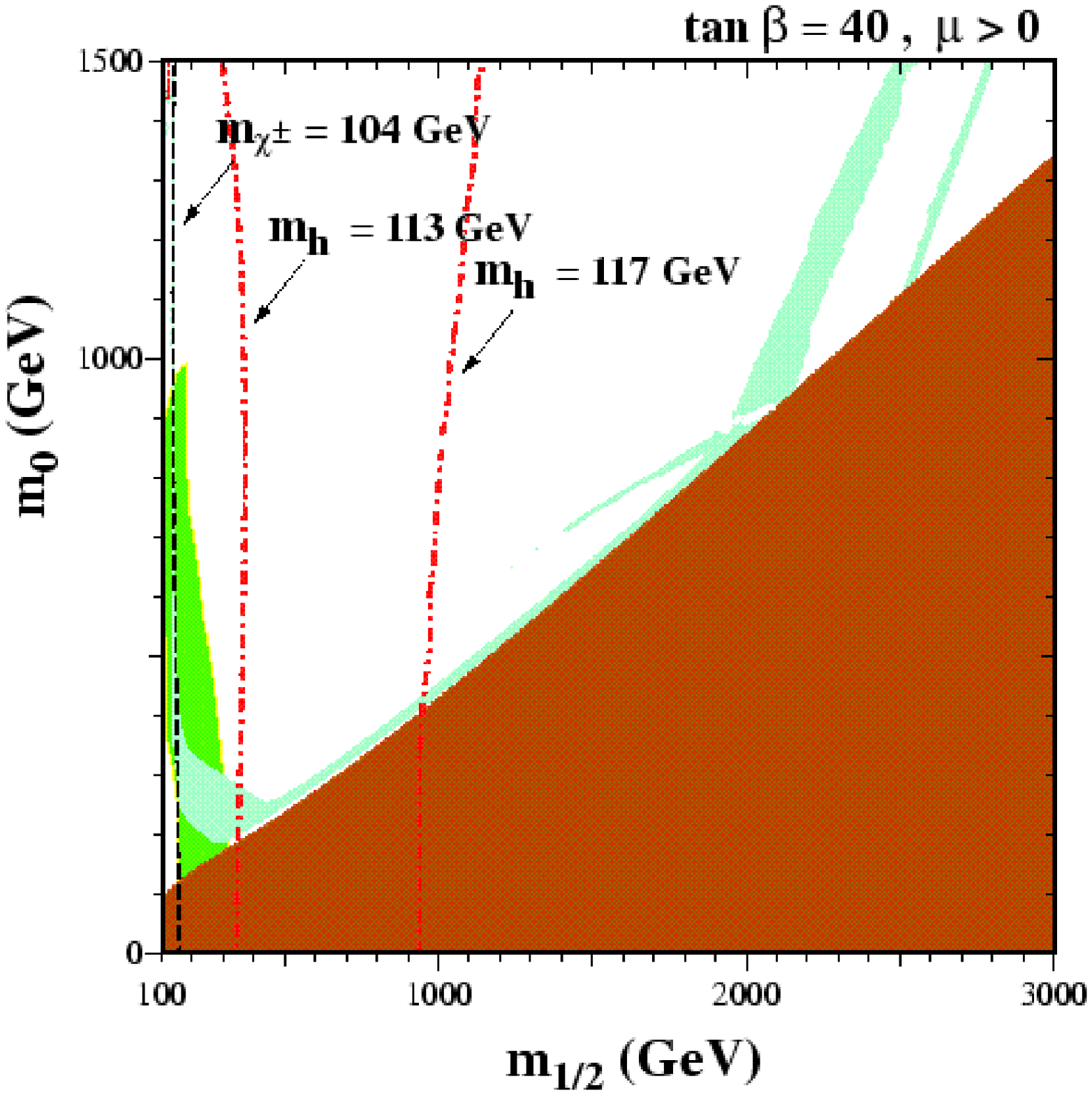,height=3.5in} \hfill
\end{minipage}
\hspace*{-.70in}
\begin{minipage}{8in}
\epsfig{file=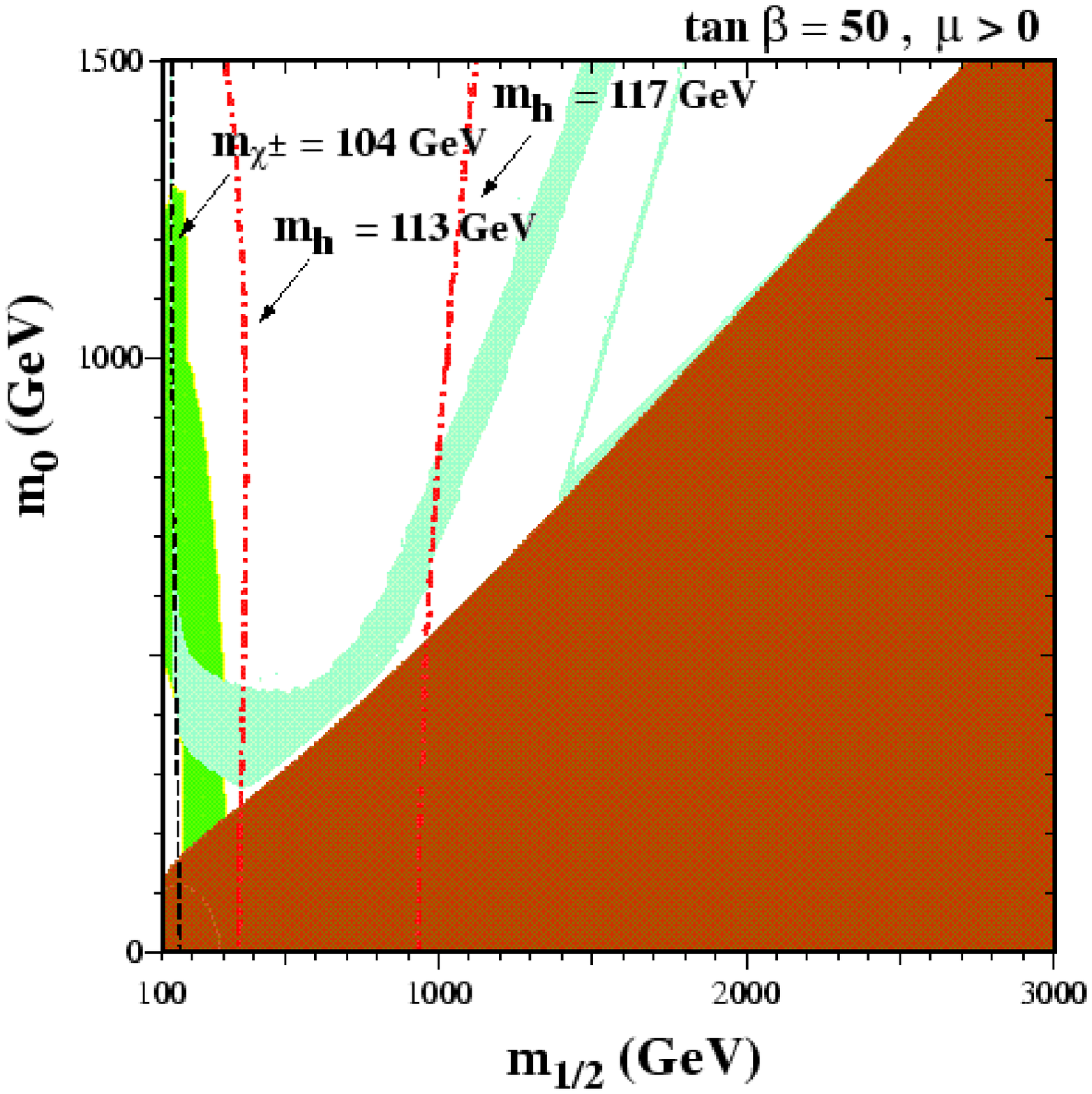,height=3.5in}
\hspace*{-0.1in}
\epsfig{file=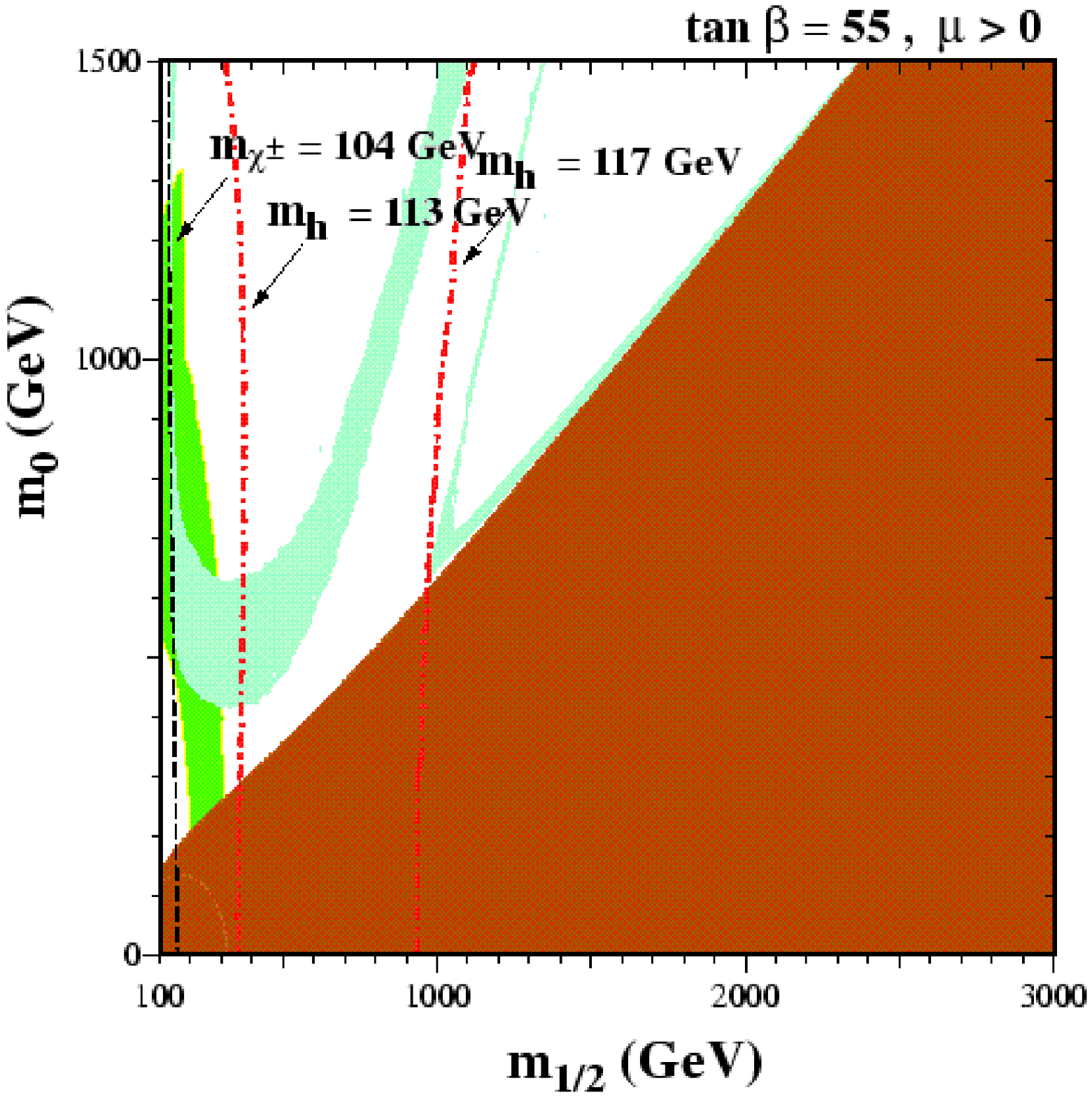,height=3.5in} \hfill
\end{minipage}
\caption{\label{fig:positive} 
{\it The $(m_{1/2}, m_0)$ planes for $\mu > 0$ and $\tan \beta =$ (a) 30,
(b) 40, (c) 50 and (d) 55, found assuming $A_0 = 0, m_t = 175$~GeV and
$m_b(m_b)^{\overline {MS}}_{SM} = 4.25$~GeV. 
The near-vertical lines are
the contours
$m_{\chi^\pm} = 104$~GeV (dashed), $m_h = 113, 117$~GeV (dot-dashed). 
The medium (dark green) shaded regions are excluded by $b \to s \gamma$.
The light (turquoise) shaded areas are the cosmologically
preferred
regions with \protect\mbox{$0.1\leq\ohsq\leq 0.3$}. In the dark
(brick red) shaded regions,
the LSP is the charged ${\tilde \tau}_1$, so this
region is excluded. 
The diagonal channel
of low relic densities visible for $\tan \beta \ge 40$, flanked on both
sides by cosmologically preferred regions, is due to direct-channel
annihilation via the $A,H$ poles.}}
\end{figure}

We display in Fig.~\ref{fig:positive} the corresponding $(m_{1/2}, m_0)$
planes for $\mu > 0$ and $\tan \beta = 30, 40, 50$ and $55$. At these
large values of $\tan \beta$, the constraint on $m_{1/2}$ from $m_h$
hardly varies as $\tan \beta$ increases, and the
region excluded by $b \to s \gamma$ is much smaller than for $\mu < 0$.
We also note that there is a small region not excluded by $b \to s
\gamma$ for $\mu > 0, m_{1/2} \sim 100$~GeV and $m_0 \la 300$~GeV, where
the conditions for the NLO treatment are not met, so that the theoretical
errors are large. However, this region is excluded by the LEP constraint
on $m_h$. 

The principal novelty in panel (a) is that the new coannihilation diagrams
and improved treatment of ${\tilde \tau}$ mixing at large $\tan \beta$
increase significantly the upper limit on $m_{1/2}$. Whereas we previously
found~\cite{EFOSi} $m_{1/2} \la 1400$~GeV when $\tan \beta \le 20$, as
also seen in panel (a) of Fig.~\ref{fig:negative}, we now see that
$m_{1/2} \la 1700$~GeV is allowed for $\tan \beta = 30$. Also,
$m_{1/2} \la 1900$~GeV is allowed for $\tan \beta = 35$, as seen
in panel (c) of Fig.~\ref{fig:negative}, and even larger
values of $m_{1/2}$ are allowed for larger values of $\tan \beta$ for
$\mu > 0$, where we find  $m_{1/2} \la 2200$~GeV for $\tan \beta \sim
50$ as seen in (c, d) of Fig.~\ref{fig:positive}.  One
effect of this extension of the cosmologically-favoured region to larger
$m_{1/2}$ is that the detection of CMSSM sparticles at
the LHC is {\it not guaranteed} for large $\tan \beta$, unlike the case
when $\tan
\beta \le 20$. At larger
$m_0$, this extension meets up with the focus point region of
\cite{Feng}. We also see in the panels (b, c) and (d) of
Fig.~\ref{fig:positive} for
$\mu > 0$ the appearance of rapid direct-channel annihilation via the
$A, H$ poles for
$\tan \beta \ge 40$. We note that this feature develops at larger $\tan
\beta$ than in the case $\mu < 0$, and recall that consistent electroweak
vacua
may readily be found for larger values of $\tan \beta$, up to
about 60. Detection
of CMSSM sparticles at the LHC is {\it also not
guaranteed} in the direct-channel $A, H$ annihilation region, whose full
extent out to large
$m_{1/2}$ and $m_0$ is not shown: it meets the focus-point
regions at very large $m_0$. This LHC-unfriendly $A, H$ annihilation 
region is rather larger than the tail of the coannihilation strip at large
$m_{1/2}$.

\begin{figure}
\vspace*{-0.75in}
\hspace*{-.70in}
\begin{minipage}{8in}
\epsfig{file=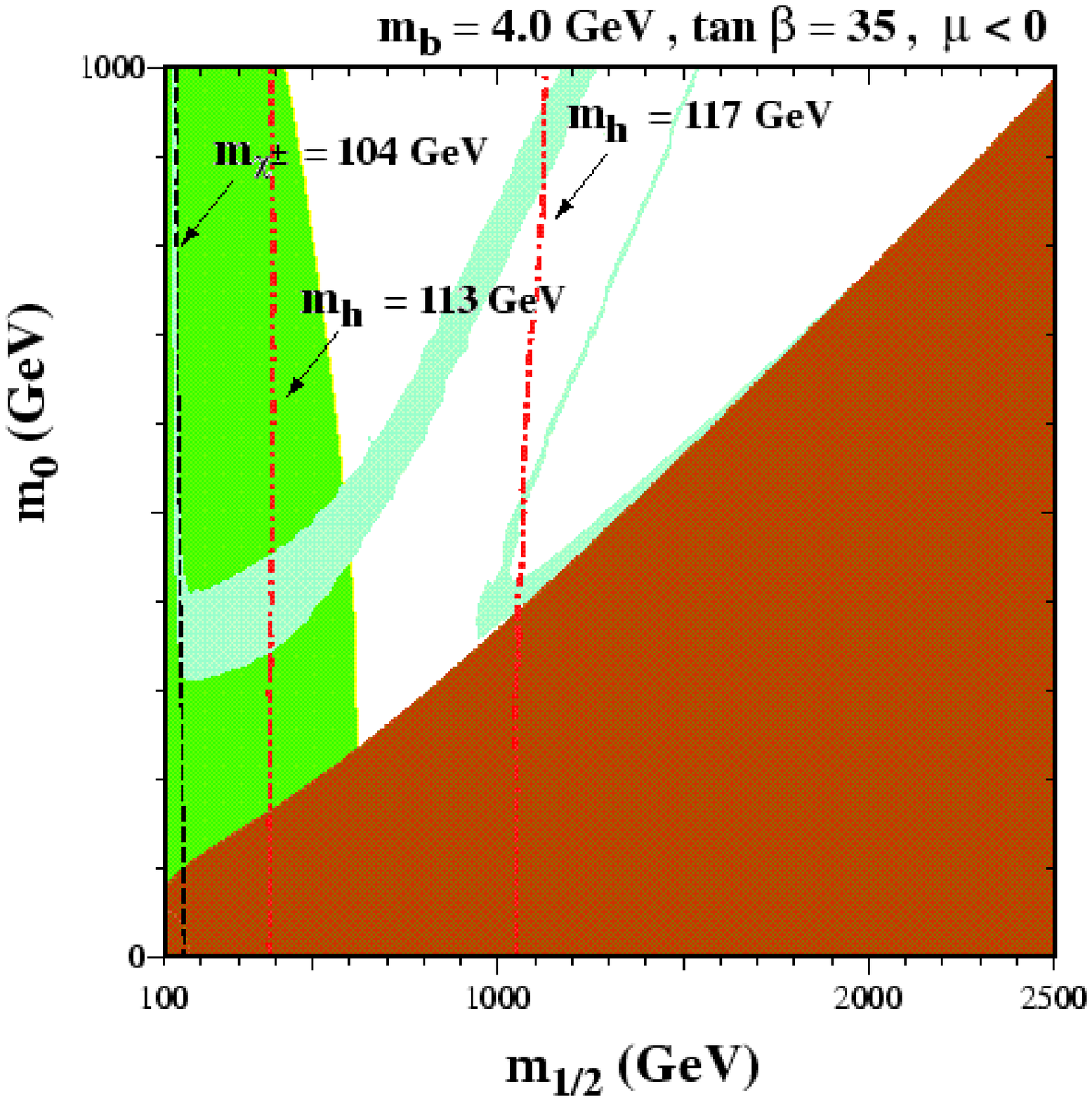,height=3.5in}
\hspace*{-0.1in}
\epsfig{file=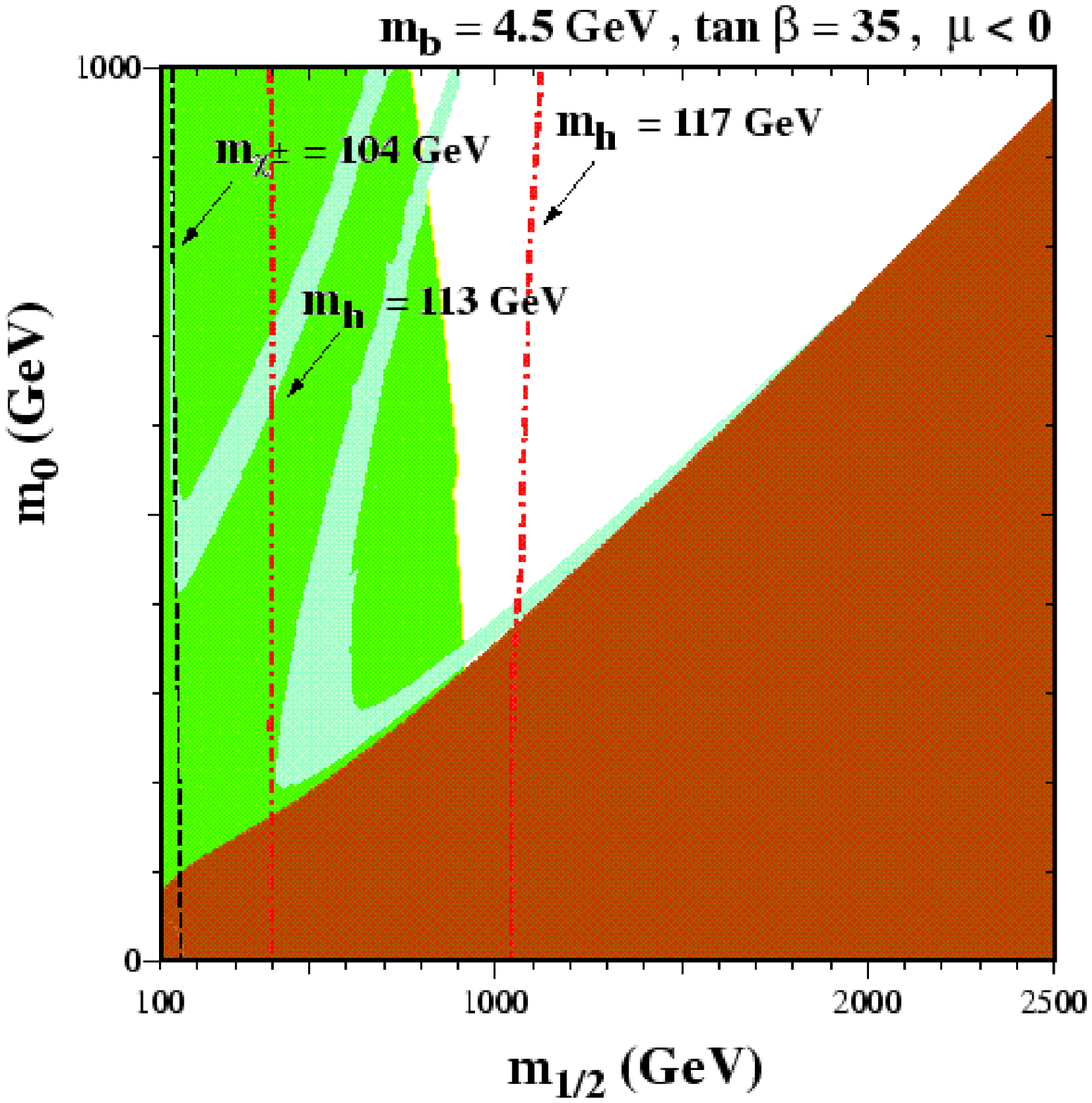,height=3.5in} \hfill
\end{minipage}
\hspace*{-.70in}
\begin{minipage}{8in}
\epsfig{file=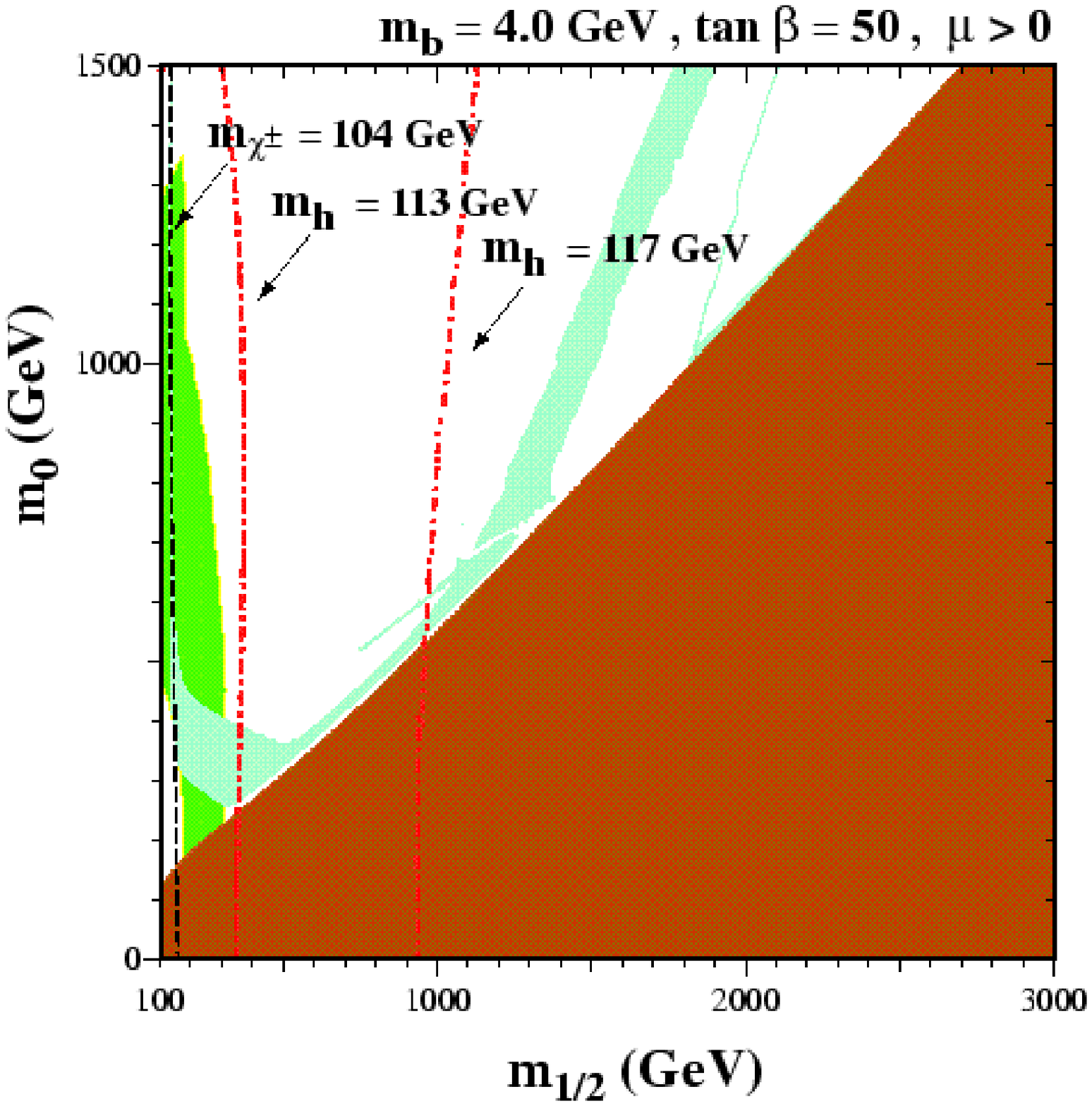,height=3.5in}
\hspace*{-0.1in}
\epsfig{file=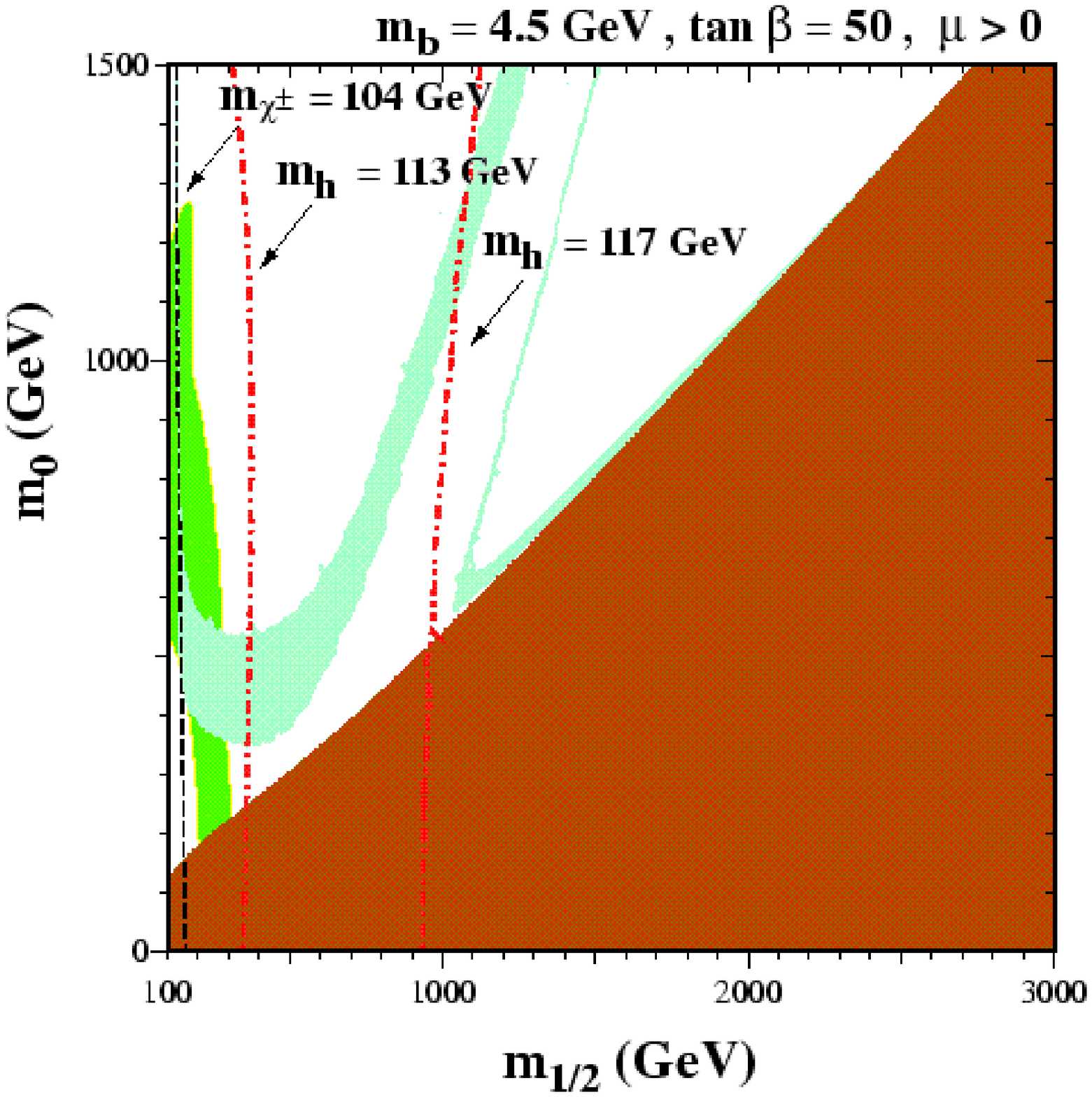,height=3.5in} \hfill
\end{minipage}
\caption{\label{fig:mbeffect}
{\it Comparison between the $(m_{1/2}, m_0)$ planes 
for $m_b(m_b)^{\overline {MS}}_{SM} 
= 4.0$ and $4.5$~GeV. Panels (a, b) are for for $\mu < 0, \tan
\beta = 35$ and panels (c, d) for $\mu > 0, \tan \beta = 50$.
In all cases, we use $m_t = 175$~GeV and  $A_0 = 0$.
We see that the channel of low relic density due to rapid
annihilation via the $A,H$ poles appears at smaller values of $m_{1/2} /
m_0$ when $m_b(m_b)^{\overline {MS}}_{SM}$ is larger.}}
\end{figure}

The sensitivity to $m_b(m_b)^{\overline {MS}}_{SM}$ at
large $\tan \beta$ is shown in Fig.~\ref{fig:mbeffect}. 
Panels (a, b) are for
$\tan \beta = 35$ and $\mu < 0$, and panels (c, d) are for $\tan \beta 
= 50$ and $\mu > 0$. 
The effect on
the LEP Higgs constraint is not
important, but the $b \to s \gamma$ constraint does depend
significantly on $m_b(m_b)^{\overline {MS}}_{SM}$. We also see in
Fig.~\ref{fig:mbeffect}
that the channel of low relic density due to rapid annihilation via the
$A,H$ poles appears at smaller values of $m_{1/2} / m_0$ when
$m_b(m_b)^{\overline {MS}}_{SM}$ is larger,
because $m_A$ decreases as $m_b(m_b)^{\overline
{MS}}_{SM}$ increases. The shift in $m_A$ is quite
significant.  Consider, for example, the point $(m_{1/2},m_0) = (900,1000)$
for $\tan \beta = 35$ and $\mu <0$, which is
along the cosmologically preferred $H,A$-annihilation strip as seen in
Fig.~\ref{fig:negative}(c).  For $m_b(m_b)^{\overline  
{MS}}_{SM} = 4.0, 4.25$, and 4.5~GeV, we find that
$m_A = 1000, 890$, and 750~GeV respectively, whereas, in each case,
$m_\chi = 404$ GeV and $m_h = 116.3$ GeV.
Furthermore, we recall that we did not find consistent electroweak vacua
when
$\mu < 0$ and
$\tan
\beta \ge 40$ for
$m_b(m_b)^{\overline {MS}}_{SM} = 4.25$~GeV, but, if we
choose $m_b(m_b)^{\overline {MS}}_{SM} = 4.0$~GeV, we find consistent
vacuum solutions for $\mu < 0$ and $\tan \beta = 40$. The
corresponding $(m_{1/2}, m_0)$ plane looks qualitatively similar to that
in panel (d) of Fig.~\ref{fig:negative} for $\mu < 0$ and $\tan \beta =
37.5$. This exemplifies the point that, although the appearance of rapid
direct-channel $A,H$-pole annihilation is a generic qualitative feature
at large $\tan \beta$, its exact location is rather model- and
parameter-dependent. 

We have also explored the sensitivity of our analysis to the assumed
values
of $m_t$ and $A_0$. The $m_h$ constraint on $m_{1/2}$ weakens
(strengthens)
significantly if $m_t = 180 (170)$~GeV~\cite{EGNO}. There are smaller
effects on the
locations of the $A,H$ poles and hence on the rapid-annihilation regions
when $\mu < 0$. For positive $\mu$, the effects are not negligible, and
for $m_t = 170$ GeV we find that the $A,H$ annihilation strip is
shifted down to lower $m_{1/2}/m_0$.  If
$A_0$ is varied, again the most significant change is that in the LEP
Higgs constraint on
$m_h$: positive (negative) values of $A_0$ increase (decrease) the Higgs
mass and weaken (strengthen) the limit, whereas changes in $A_0$ cause
only modest changes in the relic density regions. We note also that
$b \to s\gamma$ is affected slightly by $m_t$ and
to a larger extent by $A_0$. We defer detailed discussions of the
sensitivity to $m_t$ and $A_0$ to a future publication~\cite{EFGOSi2}.

\begin{figure}[htb]
\begin{center}
\mbox{\hskip -.2in \epsfig{file=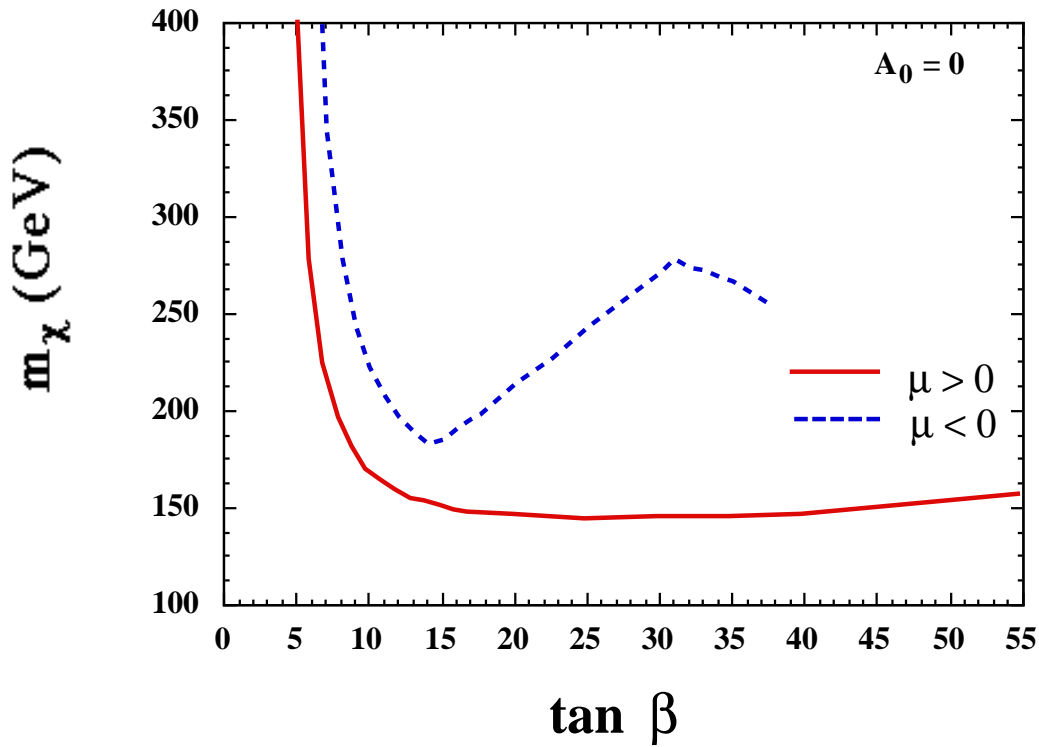,height=10cm}}
\end{center}
\caption[.]{
\label{fig:lowerlimit}
{\it Lower limits on the LSP mass $m_\chi$ obtained as functions of
$\tan \beta$ for both signs of $\mu$. We use as defaults
$m_b(m_b)^{\overline {MS}}_{SM} = 4.25$, $m_t = 175$~GeV and  $A_0 = 0$.}}
\end{figure}

Finally, we compile our results in Fig.~\ref{fig:lowerlimit} as
lower limits on the LSP mass $m_\chi$ for both signs of $\mu$ as
functions of $\tan \beta$, for our default choices
$m_b(m_b)^{\overline {MS}}_{SM} = 4.25$~GeV, $m_t = 175$~GeV and
$A_0 = 0$. The curve for $\mu > 0$ is almost the same
as in~\cite{EGNO} for $\tan \beta \le 25$.
The limits are slightly stronger here, due to the improved treatments of
the bottom quark and pseudoscalar mass. The curve for $\mu < 0$ is
resembles that in~\cite{EGNO} for
$\tan \beta \la 15$, but deviates at larger $\tan \beta$ because here we
implement the latest $b \to s \gamma$ constraint~\cite{newbsgcalx}. The
monotonic rise in the lower limit on $m_\chi$ for $15 \la \tan \beta
\la 30$ is due to the strengthening of this constraint on $m_{1/2}$ as
seen in Fig.~\ref{fig:negative}. The break and subsequent decrease in
the lower limit on $m_\chi$ at $\tan \beta \ga 30$ arise from the
intersection of the rapid-annihilation region with the weakening $b \to
s \gamma$ constraint at progressively larger $m_0$, as also seen in
Fig.~\ref{fig:negative}. As previously mentioned, we find no substantial
allowed regions of CMSSM parameter space above $\tan \beta = 37.5$ for our
default values of $m_b(m_b)^{\overline {MS}}_{SM}, m_t$ and $A_0$.
With these default values, we find $m_\chi \ga 140$~GeV for $\mu > 0$,
attained in a broad minimum around $\tan \beta \sim 25$, and $m_\chi \ga
180$~GeV for $\mu < 0$, attained for $\tan \beta \sim 15$.

A complete discussion of the absolute lower limit on $m_\chi$ as the
auxiliary parameters $m_b(m_b)^{\overline {MS}}_{SM}, m_t$ and $A_0$ are
varied over their allowed ranges lies beyond the scope of this paper, and
will be presented elsewhere\cite{EFGOSi2}. Here, we limit ourselves that
the lower limits in Fig.~\ref{fig:lowerlimit} may be reduced significantly
for different choices of these defaults. For example, if $m_t = 180$~GeV,
we find $m_\chi \ga 105$~GeV for $\mu > 0$ and $\tan \beta \sim 25$, and
$m_\chi \ga 145$~GeV for $\mu <0$ and $\tan \beta \sim 10$ (the minimum
due to the competition between the Higgs limit and the $b \to s \gamma$
constraint occurs at lower $\tan \beta$ than if $m_t = 175$~GeV). 
Similarly, if we had chosen $A_0 = 2 m_{1/2}$, as in the case of higher
$m_t$, the calculated Higgs mass, $m_h$, would be increased and the limit
on $m_{1/2}$ (and hence $m_\chi$) would be softened. In this case, we find
$m_\chi \ga 95$~GeV for $\mu > 0$ and $\tan \beta \sim 25$ and $m_\chi \ga
140$~GeV for $\mu < 0$ and $\tan \beta \sim 10$. 

As also discussed in~\cite{EFGOSi2},
these lower limits could also be relaxed somewhat if (a) $m_h$ is below
the LEP `signal' at 115~GeV and closer to the LEP lower limit of
113.5~GeV, and (b) if the theoretical calculations significantly
underestimate $m_h$, and (c) if $\mu > 0$, as can be seen in
Fig.~\ref{fig:positive}. On the other hand, if $\mu < 0$, the $b \to s
\gamma$ constraint is stronger than the Higgs constraint. If one believes
the LEP `signal', there is also an upper limit on $m_h$, and hence also on
$m_{1/2}$ and $m_\chi$. This is $m_\chi \la 400$~GeV for large $\tan
\beta$, increasing to $\sim 550$~GeV for $\tan \beta = 20$ and $\mu <
0$.

To conclude:
we have seen that an adequate treatment of the allowed CMSSM parameter
space at large $\tan \beta$ necessitates a careful analysis of
coannihilations and direct-channel $A, H$ pole annihilations. It is also
essential to treat carefully the value and the renormalization of $m_b$.
The most recent LEP lower limit on $m_h$ and recent improvements in the
experimental value and the theoretical treatment of $b \to s \gamma$ decay
at large $\tan \beta$ also play important roles. At large $\tan \beta$,
the CMSSM survives these
and the supersymmetric dark matter constraint for $\mu < 0$ only thanks to
the extensions of parameter space made by coannihilations and
direct-channel $A, H$ pole annihilations. These also play important roles
for $\mu > 0$. Putting together all the available constraints, we find
that $m_\chi \ga 140 (180)$~GeV for $\mu > 0 (\mu < 0)$ and our default
choices of $m_b(m_b)^{\overline {MS}}_{SM}, m_t$ and $A_0$. More complete
discussions of the roles of $m_t$ and $A_0$ in the CMSSM, and a treatment
of the more general MSSM without universality assumptions, are left to a
future paper~\cite{EFGOSi2}.

\vskip 0.5in
\vbox{
\noindent{ {\bf Acknowledgments} } \\
\noindent  
We would like to thank W. de Boer and P. Gambino for useful discussions.
The work of 
K.A.O. was partially supported by DOE grant DE--FG02--94ER--40823, and
that of M.S. by NSF grant PHY-97-22022.}

\end{document}